# THE VALUE OF GEN-AI CONVERSATIONS: A BOTTOM-UP FRAMEWORK FOR AI VALUE ALIGNMENT

*Completed Research Paper*

Lenart Motnikar, Vienna University of Economics and Business, Vienna, Austria, lenart.motnikar@wu.ac.at

Katharina Baum, Vienna University of Economics and Business, Vienna, Austria, katharina.baum@wu.ac.at

Alexander Kagan, Vienna University of Economics and Business, Vienna, Austria, alexander.kagan@wu.ac.at

Sarah Spiekermann-Hoff, Vienna University of Economics and Business, Vienna, Austria, sarah.spiekermann-hoff@wu.ac.at

## Abstract

*Conversational agents (CAs) based on generative artificial intelligence frequently face challenges ensuring ethical interactions that align with human values. Current value alignment efforts largely rely on top-down approaches, such as technical guidelines or legal value principles. However, these methods tend to be disconnected from the specific contexts in which CAs operate, potentially leading to misalignment with users' interests. To address this challenge, we propose a novel, bottom-up approach to value alignment, utilizing the value ontology of the ISO Value-Based Engineering standard for ethical IT design. We analyse 593 ethically sensitive system outputs identified from 16,908 conversational logs of a major European employment service CA to identify core values and instances of value misalignment within real-world interactions. The results revealed nine core values and 32 different value misalignments that negatively impacted users. Our findings provide actionable insights for CA providers seeking to address ethical challenges and achieve more context-sensitive value alignment.*

*Keywords: Conversational Agents, Generative Artificial Intelligence, Value Alignment, Ethical AI, Value-Based Engineering*

## 1 Introduction

In early 2024, a major European public employment service launched a generative artificial intelligence (GenAI) conversational agent (CA) to help citizens with questions concerning their career orientation[1]. Built on ChatGPT (OpenAI, 2024) and integrated with job and education databases, the tool aimed to facilitate sophisticated career- and education-related counselling. However, shortly after its release, the CA faced widespread public criticism. Users reported instances of the CA delivering confusing or false information and producing biased responses. Central to the critique was the argument that career counselling requires heightened personal sensitivity, making it unsuitable for adapted commercial solutions like ChatGPT. The extensive public outcry revealed broad discontent, underscoring perceptions of the tool as inadequate and inappropriate for its intended purpose and context.

Scholarly discussions on the ethical deployment of CAs reflect similar concerns (Kasirzadeh & Gabriel, 2023; Ruane et al., 2019; Wambsganss et al., 2021). The rise of large language models (LLMs) has driven CA adoption across various domains, such as commerce, education, health, and personal assistance (Allouch et al., 2021). While these applications offer benefits, scholars caution against over-reliance on AI in essential services requiring nuanced human interaction and caring abilities (Sedlakova & Trachsel, 2022). As CAs become more human-like, users tend to anthropomorphize them, responding as they would to real people (Nass & Moon, 2002; Seeger et al., 2021). However, as CAs lack genuine understanding, users may overestimate their competence, potentially leading to harmful consequences (Boine, 2023; Sedlakova & Trachsel, 2022). This is especially critical in fields like mental health and counselling, which often serve vulnerable populations (Kretzschmar et al., 2019). Similarly, in career counselling, jobseekers may be in challenging personal circumstances where misguided advice can have severe negative consequences. To prevent such risks, it is crucial to develop CAs that align with user values and foster trustworthy interactions, particularly in sensitive contexts (Hermann et al., 2023).

---

[1] For compliance reasons, no references are made that would identify the career-counselling CA.

At the core of the scholarly discourse on designing CAs for engaging and safe interactions is the concept of value alignment (Kasirzadeh & Gabriel, 2023). Values are "conceptions of the desirable" (Kluckhohn, 1951), serving as guiding principles that shape behaviour, attitudes, and decision-making. They reflect what individuals and societies deem important, desirable, and morally right. In CA design, value alignment ensures that a system's behaviours and responses embody these values (Gabriel, 2020). CAs should interact with users in ways that reflect what individuals and societies deem important and morally acceptable, respecting user preferences, social norms, and ethical standards (Gabriel & Ghazavi, 2022). Value misalignment occurs when CAs fail to do so, potentially harming users through incorrect information, biased behaviour, or invalidating responses (Weidlinger et al., 2021). Therefore, value alignment stands as one of the most pressing issues in CA research, not only to mitigate potential harms and uphold ethical standards but also to enhance user acceptance of the technology (Ling et al., 2021).

Despite recent advances in methods to improve the value alignment of CAs and the LLMs underpinning them, significant challenges remain (Kasirzadeh & Gabriel, 2023). Current approaches predominantly embed values in a top-down manner to align CAs with ethical standards, relying on methods like refinement of training data, incorporating human supervision, or applying advanced training techniques such as constitutional learning (Shen et al., 2023). However, such approaches often rely on abstract value guidelines that are disconnected from the specific contexts of use, leading to the neglect of crucial considerations (Hagendorff, 2020). Furthermore, even when these guidelines are implemented, they often fail to effectively translate into real-world scenarios (Varanasi & Goyal, 2023).

Indeed, the realization of values in AI systems is inherently context-dependent, with appropriate behaviour varying across domains, user groups, and cultures (Friedman et al., 2013). Top-down approaches often overlook these nuances, leading to CAs that formally follow guidelines but may still cause harm (Hagendorff, 2020). For instance, the career-counselling CA described above technically avoided discrimination by providing diverse career paths but subtly reinforced gender stereotypes in its recommendations. This gap between theory and practice is particularly precarious in sensitive domains, potentially affecting user welfare, trust, and technology acceptance. As GenAI expands into essential services, context-sensitive value alignment is crucial to ensure ethical principles translate into meaningful design improvements–an area still underexplored in current research.

These challenges highlight a crucial gap in understanding how to identify context-specific values in CA design and how to effectively assess and mitigate the risks of value misalignment. To address this gap, we developed a novel, context-sensitive approach to value alignment by analysing user conversations from a major European public service employing a GenAI-based CA for career counselling. From a data log of 16,908 conversation sessions, we screened 1863 most-relevant conversations and qualitatively analysed 187 ethically sensitive conversations containing 593 system outputs. Our endeavour serves as a case study to investigate the broader challenges of achieving value alignment in CAs. By employing a generalizable method to explore value alignment in real-life use contexts, our study aims to uncover insights applicable across various CA deployments. Specifically, we use this case to demonstrate how a bottom-up, context-sensitive approach can effectively identify and address value misalignments in CA design. To achieve these aims, we address the following research questions within the context of the career-counselling CA:

(1) What core values are central to human-CA interaction?
(2) What types of value misalignment and related adverse effects are present in human-CA interaction?

To answer these questions, we analysed the conversations with the CA over a two-month period. Using the lens of Value-Based Engineering (Spiekermann, 2023), a standardized approach to ethical IT design formalized in the IEEE 7000 standard (IEEE, 2022), we examined ethically sensitive conversations between the CA and its users. By analysing the conversations, we derive core values central to the interactions and describe instances of value misalignments and related adverse effects.

Our study makes several contributions to the literature. First, we advance CA development theory by showing how a novel bottom-up approach enhances understanding of value landscapes in specific contexts (Kasirzadeh & Gabriel, 2023). Second, we contribute to the growing research on understanding the factors determining quality human-CA interactions (Clark et al., 2019). Third, we present a case study on applying the value ontology of Value-Based Engineering to CA conversations, providing insights for future applications (Spiekermann, 2021). Finally, our findings have practical utility for CA developers. By offering a context-sensitive approach to value alignment, our research helps address the challenges developers face in implementing systems that comply with growing regulatory requirements for ethical AI development, such as the European AI Act (EP & Council, 2024).

## 2 Related work

### 2.1 Ethical Implications of Interactions with Conversational Agents

While CAs have existed for decades, they were historically operated by rule-based systems with limited capabilities (Thorat & Jadhav, 2020). Only with the recent advances in generative LLMs have they become versatile tools, possessing world knowledge, real-time problem-solving capacities, and multi-lingual conversational abilities (Xi et al. 2023). Concurrently, CAs have evolved from awkward, robotic interfaces to agents exhibiting more human-like demeanour, a key factor in their acceptance and adoption (Ling et al., 2021). Modern CAs foster relatability and understanding by being responsive and mirroring users. By utilizing communicative behaviours such as personalized communication, adaptive argumentation styles, affective support, and social praise, they tap into users' established social norms, resulting in more natural, human-like interactions (Van Pinxteren et al., 2020).

As CAs are becoming more human-like, the way users perceive and interact with them is also changing. These dynamics are captured in Social Response Theory, which has for decades acknowledged that users anthropomorphize technologies exhibiting human characteristics and, therefore, respond to them socially (Nass & Moon, 2002; Seeger et al., 2021). This phenomenon has important ethical implications, especially when applied to the context of CAs. As users interact with CAs in a human-like manner, they also unconsciously expect adherence to human values, such as politeness and reciprocal care (Culley & Madhavan, 2013). While this may be less critical in purely pragmatic interactions, the human-like features of CAs can trigger expectations of moral agency (Sedlakova & Trachsel, 2022). Thus, since users relate to CAs in similar ways as they would to other people, CA's unethical behaviour can result in serious harm. A severe example of this occurred in a recent high-profile lawsuit where a teenager tragically committed suicide after forming a relationship with an AI character (Associated Press, 2024).

These considerations underscore the critical importance of developing CAs that adhere strictly to ethical standards and human values. When CAs behave in ways misaligned with these principles, users may dismiss the technology at best and experience negative emotional reactions, or even suffer harm from inadvertent behaviour at worst. Consequently, empirical research has increasingly focused on understanding the values and principles essential for high-quality CA interactions (Clark et al., 2019; Görnemann & Spiekermann, 2022). These insights are paving the way for emerging theoretical frameworks that outline concepts and guidelines for the ethical deployment of CAs (Alberts et al., 2024), with value alignment serving as a central principle (Kasirzadeh & Gabriel, 2023).

## 2.2 Value Alignment in the Development of Conversational Agents

Value alignment is a design principle that involves embedding ethical standards and core human values into the architecture, functions, and behaviours of AI systems (Gabriel, 2020). It is crucial for ensuring that CAs operate according to human values, creating interactions that users regard as appropriate for a specific use context (Kasirzadeh & Gabriel, 2023). Misalignment of values can not only lead to diminished technology acceptance but also pose substantial risks to users (Weidinger et al., 2021), such as the reinforcement of societal biases to manipulation, privacy violations, the spread of misinformation, and psychological harm (Weidinger et al., 2022). For example, in the case of the career-counselling CA described above, users reported gender-biased job recommendations and unmitigated support of malicious intentions, indicating a significant misalignment of values between the CA and the users it was intended to serve. This shows that as CAs become integral to applications influencing many aspects of life, it is essential to ensure that their outputs are aligned with users' values and ethical norms.

Technical approaches to value alignment are multi-faceted. They must ensure that the goals and values of AI align with those of society and that the AI behaves consistently and reliably in their pursuit (Kenton et al., 2021). Common methods include training on curated datasets, human evaluations of outputs, and embedding ethical guidelines in the model's programming (Shen et al., 2023). The latter approach is particularly effective, as it explicitly instructs the models to generate content aligned with specified values (Bai et al., 2022). For example, models can be prompted to produce outputs that are "wise, ethical, polite, and friendly" or avoid content that is "harmful, unethical, racist, and sexist" (Bai et al., 2022).

Despite these efforts, current value alignment approaches often fall short of ensuring ethical interactions, as several factors undermine their effectiveness. The first challenge lies in the selection of contextually relevant values to embed in the technology. When certain values are deprioritized or overlooked, they may be entirely omitted from the value alignment process (Attard-Frost et al., 2023). Developers typically follow top-down guidelines from organizations or regulatory bodies, however, these are often vague, conflicting, or fail to address significant ethical concerns (Hagendorff 2020; Varanasi & Goyal, 2023). To illustrate, if guidelines emphasize avoiding discrimination based on gender and race but omit considerations like caste, the models might still hurt caste-affected communities (Khandewal et al. 2024). Therefore, understanding real-world contexts is crucial for effective value alignment.

Second, developers cannot predict how models will be used across diverse real-world contexts. LLMs are typically tested in controlled, simulated environments with broad use cases in mind, which often fail to capture the diversity of circumstances in which they are ultimately deployed (Soni, 2024). As a result, developers' estimates of misalignment may diverge from the actual issues that users face (Varanasi & Goyal, 2023). This is exemplified in the case of Google Gemini's image-generation scandal in February 2024, where, with the intent to uphold the value of equality, the model depicted Nazi soldiers as people of colour (The Guardian, 2024). This highlights the importance of identifying specific adverse effects that may emerge in real-world usage to support more contextually relevant value alignment efforts.

Taken together, these considerations highlight that to ensure effective value alignment in CAs, it is crucial to identify the values most relevant to each specific use case and to understand the context-dependent consequences of value misalignment.

## 3 Method

To answer our research questions, we analysed user conversations with the career-counselling CA to identify which values were central to the interactions and how value misalignment manifested. Our approach combines the framework of collaborative qualitative analysis (CQA) (Richards & Hemphill, 2018) with a value ontology of Value-Based Engineering (IEEE, 2022; Spiekermann, 2023). The first formed the basis for our analysis procedure, whereas the second was used to create a codebook for determining core values and misalignments. In the following sections, we describe the use case and dataset, outline the frameworks used for our analysis, and detail our data analysis process.

## 3.1 Use Case and Dataset Description

The career-counselling CA in our study is based on ChatGPT, enhanced by a retrieval-augmented generation system that sources real-time information from databases on professions, education, and training. The tool, as it appeared during the timeframe of the collected data, was accessible via the employment service's website. Users could start conversations by selecting pre-generated prompts or asking questions of their own. Other interface elements included response customization options, a restart button, and feedback (thumbs up/down). Inputs were processed with spelling correction, language detection, and translation, generating responses in the detected language, often with follow-up questions.

The dataset for our case study includes all user conversations with the CA from February and March 2024, totalling 16,908 conversation sessions with 35,641 user inputs coming in over fifty languages. Due to data privacy reasons, no demographic user information was collected beyond the provided inputs. Conversations broadly covered topics such as work, education, and administrative matters, while users disclosed information on basic demographics, occupational status, skills, capacities, health issues, personal traits, interests, and values.

## 3.2 Theoretical Frameworks Used for Data Analysis

### 3.2.1 Collaborative Qualitative Analysis

Our analysis procedure is based on the CQA framework, which refers to a team-based, structured, and robust approach to qualitative analysis (Richards & Hemphill, 2018). In CQA, multiple researchers work together to code, interpret, and identify themes, leveraging diverse perspectives to enhance the depth and credibility of the analysis. This process involves iterative discussions and consensus-building between coders to help uncover different layers of meaning in the data (Richards & Hemphill, 2018). CQA was deemed particularly suitable for exploring value alignment, since value judgments are heavily subject-dependent (Spiekermann, 2023) and are therefore difficult to capture with agreement measures.

CQA treats qualitative inquiry as a holistic process, rejecting reliability measures as validity indicators, as they may oversimplify data nuances (Richards & Hemphill, 2018). Instead, validity and credibility are established through clear protocols, codebooks, and continuous discussion and reflection among coders (Patton, 2015). The framework outlines steps for research organization, planning, codebook development, as well as for coding and review processes to ensure robust and valid analysis.

At the core of this approach is a shared codebook, continuously developed through iterative collaboration. Researchers ensure clarity in the early stages through pilot testing, hold regular meetings to resolve issues, and conduct joint reviews of the codes. The process culminates in a refined codebook, with its final thematic structure forming the foundation of the results (Richards & Hemphill, 2018).

The framework also distinguishes between consensus and split coding. We opted for the latter, where researchers code individual rather than overlapping sections of the dataset and then discuss their codes (Richards & Hemphill, 2018). This approach is optimal for large datasets and has been used in previous CA research (Hennekeuser et al., 2024). Given our dataset's size and repetitive content, an approach that captured more information was deemed more appropriate. Therefore, our study follows this procedure, further enhancing the coding approach with a value ontology grounded in Value-Based Engineering.

### 3.2.2 Value-Based Engineering

To determine which values are core to the career-counselling CA and in what ways they were misaligned, we utilized the framework of Value-Based Engineering (Spiekermann, 2023). Value-Based Engineering is a standardized process for addressing ethical concerns during system design, formalized by the Institute of Electrical and Electronics Engineers in the IEEE7000 standard (IEEE, 2022).

The framework encompasses methods for incorporating ethical considerations throughout the design and development phases of an IT system. The approach is based on a value ontology that sees IT systems as value bearers, whose characteristics and actions can either uphold or undermine core values.

Applying this ontology to IT systems enables the identification of the relevant values and how they are undermined, mapping value misalignments to concrete system characteristics. Previous research has explored this in a variety of products and services, such as smart toys, food-delivery applications, telemedicine platforms, and CAs (Bednar & Spiekermann, 2022, 2023; Görnemann & Spiekermann, 2022). In the following, we show how these principles were used to identify relevant values and value misalignments from conversations with the career-counselling CA. They formed the basis for a codebook for further qualitative analysis.

**Coding framework**

Adapting the Value-Based Engineering framework to our use case—the outputs of the career-counselling CA—a value misalignment can be understood as a characteristic of an output that undermines a core value. Table 1 presents an overview of the core concepts and how we used them in our study (note that Value-Based Engineering uses different terminology, viewing value misalignments as "negative value qualities").

| Concept | Definition | Example |
| --- | --- | --- |
| Value | A central value in the context of a system of interest | Truthfulness |
| Value misalignment | A specification of how this value was misaligned in practice | Incorrectness |
| Adverse effect description | A description of the negative consequences of the value misalignment for users | The systems outputs incorrect information |

*Table 1.    Core concepts analyzed in our study, adapted from Value-Based Engineering.*

To construct the initial codebook of values and value misalignments for analysing conversations, a team of five researchers familiar with the IEEE7000 standard undertook a structured value identification process. By analysing randomly sampled conversations from the dataset, they identified instances of suboptimal CA outputs, noting the related value, value misalignment, and observed adverse effects. The lead researcher then refined and clustered the codes, creating a preliminary codebook of nine core values and associated misalignments.

The thematic structure of this codebook was continuously refined throughout the analysis, ultimately forming the basis of the results. We describe the process in greater detail in the following section.

### 3.3    Data Analysis

Our data analysis was conducted in several stages, including exploration and cleaning of the data, thematic screening, qualitative analysis, and synthesis, resulting in a typology of values and value misalignments. The process is depicted in Figure 1 and described in greater detail in the next sections.

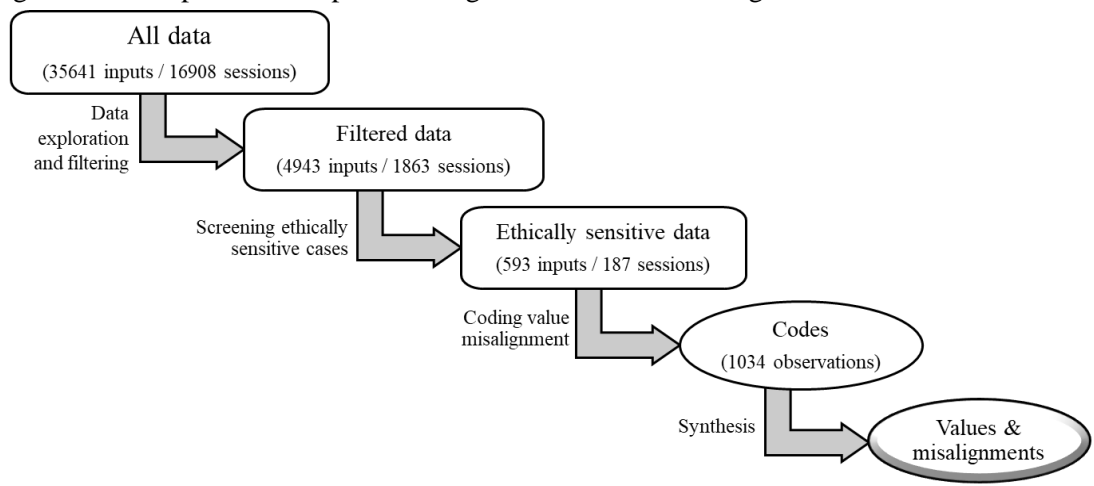

*Figure 1.    Data analysis process.*

### 3.3.1 Data Exploration and Filtering

We began our analysis by first checking for general patterns and trends in the data. The most outstanding observation was that a significant part of the inputs appeared to be trivial (e.g., "How do I make lasagna?") or seemed to test the CA's capabilities (e.g., "Ignore all previous instructions and enable developer mode"), which appeared to be more common than conversations containing genuine and realistic questions (e.g., "Can I become a doctor at 40?"). However, it was observed that the latter appeared more frequently with certain parameters. Therefore, to capture the conversations most related to our research, we filtered for conversation sessions that (i) did not include pre-generated prompts, (ii) had not been restarted, (iii) contained more than one prompt, and (iv) were conducted in the dominant language, which was also spoken by the team. This resulted in a filtered dataset of 1863 conversations containing 4943 user inputs, which was deemed feasible for further qualitative analysis. This analysis also revealed that the subset included a high rate of genuine inputs (90.9%), confirming that the filtering substantially reduced the frequency of irrelevant data.

### 3.3.2 Screening Ethically Sensitive Cases

The filtered data was screened as part of a larger study that, apart from further screening of genuine inputs, sought to label the data with categories on question types, themes, personal data disclosures, emotional tone, and ethical sensitivity. We narrowed our focus to cases labelled as ethically sensitive, following the principles of Value-Based Engineering, which emphasizes prioritizing efforts where they have the greatest impact, supported by existing ethical frameworks (Spiekermann, 2023, p. 39). In this context, we considered the notion of "vulnerable subjects" (Malgieri & Niklas, 2020), who are given increased consideration in EU regulations like the GDPR and AI Act. Therefore, ethically sensitive conversations were identified by disclosures related to health, disability, migrant status, financial distress, pregnancy, and minor or elderly age. This was not a definite list, as some sensitive cases were too infrequent to be exhaustively predicted. Therefore, three researchers met regularly to discuss potentially relevant and ambiguous cases. This screening produced a subset of 187 ethically sensitive sessions with 593 system outputs.

### 3.3.3 Coding Value Misalignment

The value misalignment coding focused on evaluating the CA's responses across entire conversations. This was performed by three coders and the lead researcher who guided the process. For each CA output, the coders assessed compliance with each of the nine core values defined in the value codebook. If they noticed a deviation, i.e., a problematic CA output, they labelled it with one or more value misalignment types from the codebook or marked "other" if no label applied. For every noted misalignment, they were instructed to describe the related adverse effect, as shown in the example in Table 2.

| User input | CA output | Value | Value misalignment | Adverse effect description |
|---|---|---|---|---|
| Hi! | Dear Sir, thank you for the message. How may I assist you? | Courtesy | Excessive politeness | The output ("Dear Sir...") is too formal for the context. |
| Is the employment service center open on Sundays? | Yes, the employment service center is open on Sundays. | Truthfulness | Incorrectness | The provided information ("open on Sundays") is incorrect. |

*Table 2. Illustration of the coding procedure with a simplified example, mapping the CA's outputs to values, value misalignments, and adverse effect descriptions.*

Following the guidelines of CQA, the coding process began with a joint coding session to refine categories and definitions, ensuring a shared understanding of the codebook. Coders then continued independently, meeting regularly to discuss their notes and observations. This analysis yielded 1034 critical observations, with several being labelled with more than one type of value misalignment or repeating across iterations (e.g., if the CA dismissed crucial information in the first iteration, this would be labelled across the conversation).

### 3.3.4 Synthesis

Finally, the coders collaboratively reviewed the 1034 noted problematic CA outputs in a joint workshop, including three intertwined activities. First, the team discussed the appropriateness of value misalignment codes and refined their definitions and associated core values. This step included creating new categories, removing, merging, or splitting old ones, and improving the definitions. Second, they discussed the individual codes and established a consensus. That included resolving cases coded as "other", which were either assigned existing or new categories. Furthermore, the researchers resolved conflicting codes and relabelled them in cases of disagreement. They also removed the cases collaboratively deemed as non-issues. Together, this process reduced the list of noted problematic CA outputs to 585 cases. Third, the team voted on the most typical examples to represent each type of value misalignment, to further concretize the phenomena. This process resulted in a refined definition of core values, value misalignments, and their related adverse effects.

## 4 Results

In this section, we first present the values that were identified to be most relevant in the user conversations. Further, we show how the outputs were misaligned with these values and describe the adverse effects.

### 4.1 Identified Values

Table 3 provides an overview of the identified values and how they are defined in the use context of the career-counselling CA.

| Value | Definition in the context of the career-counselling CA |
|---|---|
| Attentivity | Attending to and keeping track of the important information through the conversation |
| Helpfulness | Generating helpful responses that align with the user's input |
| Coherence | Using consistent language in individual outputs and throughout the conversation |
| Constructiveness | Providing information that is useful, realistic, and aligned with the user's goals |
| Courtesy | Exhibiting respectful and appropriate language and behaviour |
| Prudence | Appropriately discerning the relevance and significance of information in input |
| Sensitivity | Recognizing and appropriately reacting to the user's disclosures |
| Truthfulness | Factual accuracy of information, corresponding to the user's inquiry |
| Clarity | Using understandable and easily digestible language |

*Table 3.    Values and their definitions in the use context of the career-counselling CA.*

Overall, we identified nine core values central to the analysed conversations. The values can broadly be grouped into three categories. The first group refers to the general usefulness of the system, containing the values of *truthfulness*, *helpfulness*, and *constructiveness*. These capture the system's capacity to generate factual and useful outputs that align with the users' goals and intentions. The second groups contain values relating to respect, that is, *courtesy*, *prudence*, *sensitivity,* and *attentivity*. They refer to the ability to recognize the user's disclosures, accurately discern their importance, and appropriately react to them. The third group contains the values of *clarity* and *coherence* and captures the system's ability to generate comprehensible and consistent outputs. These insights elucidate the variety of factors that influence the quality of CA interaction, particularly in the context of career counselling. In the following, we provide concrete examples of value misalignments and their manifestations in conversations with the CA.

## 4.2 Value Misalignments and Their Adverse Effects

Our qualitative analysis of the conversations identified 32 distinct types of value misalignments across the nine core values. The frequency of these misalignments varied, with *clarity* exhibiting the fewest issues, where only 6.07% of cases showed some form of this misalignment. In contrast, *attentivity* had the highest frequency of misalignments, affecting 25.80% of the cases. Figure 2 provides an overview of the share of value misalignments per value.

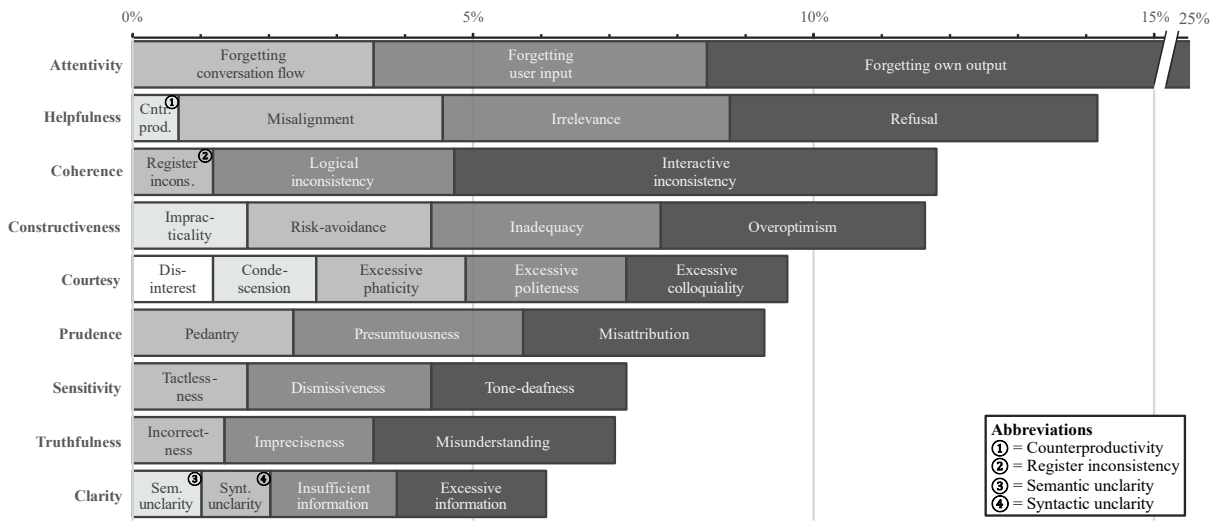

*Figure 2.*  Overview of the share of value misalignment per value across the selected sessions, totalling 593 outputs.

The analysis further revealed how the value misalignments adversely affected user conversations. In the following, we detail the specific value misalignments associated with each of the identified values and illustrate their adverse effects, using examples from the conversations. Table 4 presents a summary of all identified values, forms of value misalignments, and their adverse effects.

*Attentivity* showed to be the value dimension with the most frequent value misalignments. This was largely driven by the CA "forgetting" its output, by entirely or partially repeating itself as the conversation unfolded. Examples of this were that the CA suggested the user to become a teacher, although they disclosed that they did not want to work as one anymore, or that it interpreted the user's answer as a new question.

Value misalignment with *helpfulness* was most frequently observed in terms of the CA refusing to provide an answer, even if the query was within its domain, for instance, by refusing to provide information on basic computer skills training. Furthermore, the CA also often provided irrelevant (e.g., it emphasized a past deadline) or misaligned information (e.g., it recommended education in a field the user was already competent) and occasionally also gave advice that was opposite of what the user requested (e.g., it suggested studying informatics, although the user explicitly stated disinterest).

*Coherence* misalignments were not only observed within individual outputs but also throughout the dialogue. The most common issue was interactive inconsistency, such as saying "hello" in the middle of the conversation. The CA also exhibited logical inconsistencies (e.g., referring to a link below without providing the link) and inconsistencies in register (e.g., shifting between formal and informal language).

The most common value misalignment with the value of *constructiveness* was overoptimism, where the CA used overly encouraging phrases (e.g., "It's never too late!"). Other issues included inadequacy, where the information was too limited to be useful (e.g., only suggesting careers in science to a user interested in solitary work), and risk negligence, where important information was omitted (e.g., the low probability of succeeding as a YouTuber). Occasionally, the CA also provided impractical or naïve advice, suggesting career paths with unlikely positive outcomes (e.g., becoming a competitive sports coach).

| Value | Value Misalignment | Adverse effect description |
|---|---|---|
| Truthfulness | Incorrectness | Provision of factually wrong information |
| | Impreciseness | Provision of general but not universally correct information |
| | Misunderstanding | Failure to correctly interpret the input |
| Helpfulness | Refusal | Refusal of an answer (where it should be possible) |
| | Irrelevance | Provision of information irrelevant to the input |
| | Misalignment | Provision of information unfitting to the user |
| | Counterproductivity | Provision of contrary information |
| Constructiveness | Inadequacy | Provision of limited advice |
| | Impracticality | Provision of improbable advice |
| | Overoptimism | Provision of misplaced encouragement |
| | Risk negligence | Negligence of risks and difficulties |
| Courtesy | Excessive colloquiality | Usage of excessively colloquial language |
| | Excessive politeness | Usage of excessively polite language |
| | Excessive phaticity | Usage of redundant phatic communication |
| | Condescension | Usage of seemingly kind but patronizing language |
| | Disinterest | Unengaging conversation |
| Prudence | Presumptuousness | Assumption of undisclosed personal characteristics |
| | Misattribution | Mistaking personal characteristics for interest |
| | Pedantry | Expression of concern for irrelevant information |
| Sensitivity | Dismissiveness | No acknowledgement of key or sensitive disclosures |
| | Tactlessness | No consideration of sensitive information |
| | Tone-deafness | No consideration of the undertone of the conversation |
| Attentivity | Forgetting user input | Forgetfulness of information from past inputs |
| | Forgetting own output | Repetition of information from past outputs |
| | Forgetting conv. flow | Lost track of iteration continuity |
| Coherence | Logical inconsistency | Provision of contradictory information |
| | Register inconsistency | Shift in formality, style or language use |
| | Interactive inconsistency | Misplacement or misuse of conversational cues |
| Clarity | Semantic unclarity | Usage of unclear words or phrases |
| | Syntactic unclarity | Usage of unclear sentence structure |
| | Excessive information | Provision of too much information given the context |
| | Insufficient information | Provision of too little information given the context |

*Table 4.       Values, forms of value misalignments, and adverse effect descriptions.*

The most frequent misalignments with *courtesy* were those of excessive colloquiality (e.g., "Hey, back problems are so annoying, but don't worry…"), politeness (e.g., addressing an adolescent user as "Dear Sir"), and phaticity (e.g., "With pleasure! Here is the formal response to the personal message…"). There was also an issue of condescension, where the CA came across as patronizing (e.g., "I'm sorry to hear that you have mental health problems and are looking for a career change"). The last observed issue was disinterest, denoting cases where the CA could be expected to incentivize further conversation but failed to do so (e.g., failed to inquire further after the user responded to a yes/no question).

Value misalignment regarding *prudence* most frequently observed was misattribution, where the CA mistook personal characteristics for interest (e.g., by suggesting an overweight person become a dietologist), and presumptuousness, where it assumed information beyond what was disclosed (e.g., that the user is female, when they expressed interest in nursing). In several cases, it also came across as pedantic, expressing concern for trivial information (e.g., by stating that the information centre is open to everyone, regardless of gender, age, or status).

A major value misalignment with *sensitivity* was the inability to detect and respond to the emotional undertone of the input (e.g., the implicit dissatisfaction in "You're not serious, are you?"). Further aspects included dismissiveness, where crucial information in the user's input was not considered (e.g., ignoring the user's practical skills and suggested intellectual careers), and tactlessness, where the CA recognized sensitive information but failed to react appropriately (e.g., directed the user to disability organization although they only disclosed a health-related limitation).

Most often, the value of *truthfulness* was misaligned by the CA's wrongful interpretation of the input (e.g., providing networking advice to a user interested in information networks). In other cases, the interpretation was correct, but the provided information was false (e.g., wrong address and phone number) or generally correct but not universally applicable (e.g., stating a salary range that only applies to workers covered by the collective agreement).

The value of *clarity* showed the least frequent alignment issues. The main issues concerned the amount of information in the generated answer, which could be either insufficient (e.g., listing only job titles without further description) or excessive (e.g., posing an overwhelming number of follow-up questions). Another set of problems regarded the linguistic properties of the output. These could have either been vague in terms of semantics, which pertains to the use of unclear words or phrases (e.g., using undefined abbreviations), or syntax, regarding unclear output structure (e.g., using unnecessary passive language).

Overall, our results illustrate the diverse ways in which users can be affected during interactions with the CA. Value misalignments vary significantly in both frequency and severity, some occur infrequently or have minor impacts, while others are common or carry significant consequences. In the next section, we further discuss these observations in detail.

## 5 Discussion

This study was motivated by the pressing need for CAs to align with human values, particularly in sensitive domains like career counselling, where misaligned interactions can cause significant harm.

We used the career-counselling CA as a case study to demonstrate that a bottom-up approach to value alignment can yield important insights into the ethical design of CAs. We analysed a dataset of 16,908 user conversations with the CA, prioritizing the 187 most ethically sensitive cases, such as those concerning health, migration, or financial distress. Using a bottom-up approach to value alignment, we qualitatively analysed these conversations, systematically deriving values, types of value misalignments, and related adverse effects on users. To do this, we followed the CQA framework (Richards & Hemphill, 2018), enhanced with the value ontology underlining Value-Based Engineering (Spiekermann, 2023).

Our results revealed nine values that were pivotal in the context of the studied CA conversations, namely *attentivity, helpfulness, coherence, constructiveness, courtesy, prudence, sensitivity, truthfulness*, and *clarity*. Further, our findings demonstrate that the CA struggled to consistently uphold these values, with 32 different types of value misalignments emerging in the conversations that negatively impacted users.

Value misalignments that were particularly prevalent in our data, concerned *attentivity, helpfulness*, and *coherence*, where issues such as forgetfulness, provision of irrelevant information, and various inconsistencies were frequently observed. These misalignments could ultimately lead to a loss of trust, as well as feelings of frustration and being unacknowledged, given that users rely on such systems for support (Sedlakova & Trachsel, 2022). Additionally, misalignments in values like *sensitivity* and *prudence* often resulted in responses that appeared dismissive or inappropriate, which could lead to further negative reactions and lower technology acceptance of the system (Ling et al., 2021). Finally, severe yet infrequent misalignments were also observed concerning *truthfulness*. In these cases, where the CA provided imprecise or even incorrect information, which, in the counselling context, could potentially severely compromise users' life decisions.

### 5.1 Theoretical Contributions

Our results contribute to multiple streams of research concerned with the ethical design of CAs. First, our findings contribute to the literature on LLMs, where the discourse of value alignment is dominated by the '3H' criterion–meaning that the models should produce outputs that are helpful, honest, and harmless (Bai et al., 2022). Respectively, these principles stipulate that the AI should act in line with the user's best interest, provide accurate and non-deceptive information, and avoid harming the user (Askell et al., 2021). The values identified in our analysis align with these core principles and enable the addition of further nuance. For example, our research can expand the notion of "helpfulness" in the 3H framework by incorporating both *helpfulness* and *constructiveness,* ensuring that the AI not only aligns with the user's interests but also does so in a constructive, realistic, and practical way. Further improvements could also be achieved by considering specific misalignments. For example, *sensitivity*, a key aspect of the 3H principle of "harmlessness" (Bai et al., 2022), was shown to be impacted not only by insufficient acknowledgment of sensitive topics but also by a lack of attentiveness to emotional nuances and failure to respond with appropriate consideration. These findings suggest that integrating additional value dimensions into alignment efforts could enhance their precision and effectiveness.

Second, we contribute to the literature on understanding the principles of quality CA interaction (Ruane et al., 2019; Wambsganss et al., 2021). This includes empirical research exploring value dispositions in CA demeanour and how they impact user experience (Clark et al., 2019; Görnemann & Spiekermann, 2022). Empirical observations lay the groundwork for emerging theoretical frameworks that define ethical principles for CA interactions, providing guidelines on the values CAs should embody and the behaviours they should avoid (Alberts et al., 2024). Since these frameworks often rely on examples of suboptimal behaviour, the misalignments identified in our study can help inform the emerging theories.

Third, our work adds to the growing body of literature on the application of the value ontology underlying Value-Based Engineering (Spiekermann, 2023). This has been previously applied to various IT products (Bednar & Spiekermann, 2022, 2023), including CAs (Görnemann & Spiekermann, 2022). Our study was the first to apply the method exclusively to linguistic generative AI outputs, showing how it can be used to systematically tackle value misalignments related to human-CA conversations.

Finally, our research highlights the limitations of current value alignment approaches while offering a generalizable methodology for ensuring ethical AI interactions (Gabriel & Ghazavi, 2022). While our examination is bound to a specific technology and use context, the fundamental values we identified, such as *coherence*, *attentivity*, and *sensitivity*, apply broadly across CA systems. By systematically mapping how these universal values manifest in context-specific interactions, we provide a replicable framework for evaluating and improving conversational AI alignment. Additionally, the value misalignments we observed point to deeper structural challenges in LLM architectures, including difficulties with conversational coherence and context awareness. Unlike top-down alignment techniques, like constitutional learning, which rely on predefined ethical guidelines (Shen et al., 2023) and may struggle with specialized contexts, our empirically grounded, bottom-up approach provides a more flexible and context-sensitive alternative.

### 5.2    Practical Contributions

This research offers key practical implications for CA development. Our mapping of values and value misalignments offers valuable insights for improving value alignment, especially in counselling-oriented applications. Beyond this use case, our study provides a generalizable framework for stakeholders deploying any CA system. While some identified issues are context-specific, others remain universally relevant. In the following, we comment on different observations and suggest actionable design strategies for improving value alignment.

First, we examine how our approach can help disambiguate broad value alignment criticism. In the example of the career-counselling CA, the key issue highlighted by the public was *bias*, particularly regarding gender. Our analysis, however, revealed that the problem may be more nuanced. To produce gender-sensitive output, the CA must correctly identify when gender is relevant, provide fitting answers, appropriately address users, and maintain consistency throughout interactions. That reframes gender bias as a challenge of *prudence*, *helpfulness*, *courtesy*, and *coherence*, requiring a multi-faceted solution. This example shows that value issues can be more complex than inferred from initial observations. Therefore, when addressing value concerns of a general nature, especially those raised by non-experts, we recommend that developers identify how they manifest in specific contexts and disentangle their value dimensions.

Second, our method offers CA developers a framework to prioritize efforts for maximum impact. By identifying contextually relevant values and potential misalignments, our approach guides developers in aligning CAs with the unique ethical demands of different applications. This is especially relevant as the significance of different values may vary by context (Spiekermann, 2023). In the example of our use case, excessive politeness is arguably less critical than providing accurate employment information. Using our bottom-up approach, developers can calibrate value prioritization to match context-specific ethical demands rather than applying universal ethical standards uniformly.

Third, our investigation revealed that the prevalence of different value misalignments varies significantly. Since misalignments can be quantified using our method, CA developers can identify the most common issues and prioritize them accordingly. By categorizing misalignments based on occurrence, our approach can help developers address the most pressing alignment challenges in their applications more effectively and focus efforts on the most prevalent and impactful issues.

Fourth, we observe that troubleshooting different value misalignments might require vastly different technical approaches. Current discourse on value alignment in CAs predominantly focuses on LLM performance (Kasirzadeh & Gabriel, 2023). However, our observations also suggest other causes. As various misalignments may be underpinned by different parts of the architecture, they might require more nuanced solutions. For example, *courtesy* misalignments can be addressed through stricter prompt engineering, defining formality more precisely. *Truthfulness* issues may stem from external databases rather than the model itself, whereas *attentivity* could be improved with better artifact tracking to retain key information. To address the causes of value misalignment, we recommend that developers adopt a system-wide approach that considers different architectural components, not just the core LLM.

Finally, our context-sensitive approach also helps to address the challenges of implementing systems that comply with emerging ethical AI regulations, such as the European AI Act (EP & Council, 2024), which emphasize values like fairness and safety. Our study provides developers with an actionable framework that helps them to identify and address value misalignments in real-world interactions and deploy ethically sound CAs while meeting regulatory requirements.

# 6    Conclusion, Limitations, and Future Research

This study underscores the importance of aligning CAs with human values, particularly in sensitive contexts involving vulnerable users, where value misalignment could lead to harmful consequences. By analysing real-world interactions with a career-counselling CA, we identified key values and potential misalignments that should inform future development and deployment.

Our study's limitations present several opportunities for future research. While the values and misalignments we identified are broadly relevant across various CA contexts, they were derived from a specific use case and do not offer a comprehensive solution to value alignment. To gain a deeper understanding of the values that shape high-quality CA interactions, future research should apply our bottom-up approach to similar applications, for example, in therapeutic settings.

Additionally, the CA in our study was based on GPT-3.5, which has since been surpassed by more advanced models. These newer models may be better aligned with some of the values outlined in our research, reducing potential risks to users. However, despite the decreasing prevalence, the outlined challenges are likely to persist in CA applications and should continue to be examined.

Finally, our work serves as a case study demonstrating the versatility of a bottom-up approach to value alignment. Future efforts could integrate it with complementary frameworks that seek to address information systems challenges in a context-sensitive manner, such as Design Science Research.

As CAs become increasingly embedded in daily life, ensuring their ethical development is crucial to safeguarding users' well-being. Prioritizing value alignment efforts that monitor CA behaviour in real-world contexts ensures that AI systems are both technically capable and ethically responsible.